\documentclass[usenatbib]{mn2e}
\usepackage{graphicx,times}
\usepackage{bm}
\graphicspath{{./fig/}{./png/}}
\newcommand{\mean}[1]{\overline{#1}}
\newcommand{\meanv}[1]{\overline{\bm #1}}
\newcommand{\pd}{\partial}
\newcommand{\urms}{u_{\rm rms}}

\newcommand{\brms}{B_{\rm rms}}

\newcommand{\kef}{k_{\rm f}}
\newcommand{\Beq}{B_{\rm eq}}
\newcommand{\etat}{\eta_{\rm t}}

\newcommand{\Sh}{{\rm Sh}}

\newcommand{\Ra}{{\rm Ra}}
\newcommand{\Co}{{\rm Co}}
\newcommand{\Pm}{{\rm Pm}}
\newcommand{\Rey}{{\rm Re}}
\newcommand{\Rem}{{\rm Rm}}
\newcommand{\memf}{{\mathcal{\mean{\bm E}}}}
\newcommand{\memfy}{{\mathcal{\mean{E}}_{y}}}
\def\onehalf{{\textstyle{1\over2}}}
\def\onethird{{\textstyle{1\over3}}}
\newcommand{\Fig}[1]{Fig.~\ref{#1}}

\newcommand{\Eq}[1]{equation~(\ref{#1})}

\newcommand{\yaj}[3]{ #1, {AJ,} {#2}, #3}

\newcommand{\yapj}[3]{ #1, {ApJ,} {#2}, #3}

\newcommand{\yan}[3]{ #1, {Astron.\ Nachr.,} {#2}, #3}

\newcommand{\yznat}[3]{ #1, {Z.\ Nat.,} {#2}, #3}

\newcommand{\yana}[3]{ #1, {A\&A,} {#2}, #3}

\newcommand{\yjfm}[3]{ #1, {J.\ Fluid Mech.,} {#2}, #3}
\newcommand{\ypepi}[3]{ #1, {Phys.\ Earth Planet.\ Int.,} {#2}, #3}

\newcommand{\yprl}[3]{ #1, {Phys.\ Rev.\ Lett.,} {#2}, #3}

\newcommand{\ymn}[3]{ #1, {MNRAS,} {#2}, #3}

\newcommand{\ypre}[3]{ #1, {Phys.\ Rev.\ E,} {#2}, #3}

\newcommand{\yjour}[4]{ #1, {#2}, {#3}, #4}

\newcommand{\ybook}[3]{ #1, {#2} (#3)}

\title[The $\alpha$ effect in rotating convection with sinusoidal shear]
{The $\alpha$ effect in rotating convection with sinusoidal shear}
\author[P.\ J.\ K\"apyl\"a et al.]{P.\ J.\ K\"apyl\"a$^{1,2}$, 
M.\ J.\ Korpi$^{1,2}$, and A.\ Brandenburg$^{2,3}$\\
$^1$Observatory, T\"ahtitorninm\"aki (PO Box 14),
FI-00014 University of Helsinki, Finland\\
$^2$NORDITA, AlbaNova University Center, Roslagstullsbacken 23,
SE-10691 Stockholm, Sweden\\
$^3$Department of Astronomy, AlbaNova University Center,
Stockholm University, SE-10691 Stockholm, Sweden
}
\date{Accepted 2009 November 5.  Received 2009 October 30; in original form 2009 August 18}

\begin{document}
\maketitle

\begin{abstract}
Using three-dimensional convection simulations it is shown that
a sinusoidal variation of horizontal shear leads to a kinematic $\alpha$ effect
with a similar sinusoidal variation.
The effect exists even for weak stratification and arises owing to the
inhomogeneity of turbulence and the presence of impenetrable vertical boundaries.
This system produces large-scale magnetic fields that
also show a sinusoidal variation in the cross-stream direction.
It is argued that earlier investigations overlooked these phenomena
partly because of the use of horizontal averaging and also because
measurements of $\alpha$ using an imposed field combined with long time
averages give erroneous results.
It is demonstrated that in such cases the actual horizontally averaged
mean field becomes non-uniform.
The turbulent magnetic diffusion term resulting from such non-uniform
fields can then no longer be neglected and begins to balance the $\alpha$
effect.
\end{abstract}

\label{firstpage}

\section{Introduction}

Shear can play an important role in hydromagnetic dynamos.
This is especially true of dynamos in astrophysical bodies
that generate magnetic fields on
scales larger than the scale of the turbulent motions.
Those types of dynamos are generally referred to as large-scale dynamos.
Simulations confirm that shear can be the sole driver of dynamo action
\citep{B05,Yousef1,Yousef2,BRRK08}, but there is no consensus as to what
is the underlying mechanism for producing such large-scale fields.
In addition to shear there are also other possible mechanisms
producing large-scale magnetic fields.
One important contender is the $\alpha$ effect \citep{SKR66}, which
quantifies the effect of kinetic helicity on magnetic field generation.
It can also be the sole driver of large-scale dynamo action \citep{B01,KKB09b}.

When both shear and $\alpha$ effect act simultaneously, it becomes even
harder to identify the main drivers of large-scale dynamo action.
Although shear is generally believed to be advantageous for large-scale
dynamo action \citep[e.g.][]{Tobias09}, it is conceivable that the two effects
($\alpha$ effect and shear) suppress each other at least partially.
This is because, in the presence of stratification or other inhomogeneities, 
shear itself can produce an $\alpha$ effect \citep{RK03,RS06,KKB09a}.
Its sign depends on the relative orientation of shear and stratification.
The net $\alpha$ depends then on the pseudo scalar
$(2\bm{\Omega}+\overline{\bm{W}})\cdot\bm{g}$, where $2\bm\Omega$ and
$\overline{\bm{W}}$ are the vorticities associated with rotation and
large-scale shear flow, respectively.

The issue can be complicated even further if shear is not constant but
has a sinusoidal profile, for example \citep{BBS01,HP09}.
Sinusoidal shear profiles are commonly adopted in numerical simulations
where all boundaries are strictly periodic.
This has obvious computational advantages and is certainly easier to
implement than the so-called shearing-periodic boundary conditions
where cross-stream periodicity applies only to positions that follow
the shear flow and are thus changing with time \citep{WT88}.
In helical turbulence with shear there is the possibility of dynamo
waves that propagate perpendicular to the plane of the shear.
This is clearly borne out by simulations \citep{KB09}.
The propagation direction of the dynamo wave is proportional to the product
$H_{\rm K}\overline{\bm{W}}$, where $H_{\rm K}$ is the kinetic helicity
of the flow.
When the shear is sinusoidal, the sign of $\overline{\bm{W}}$ changes
in space, so one obtains counter-propagating dynamo waves in the two halves
of the domain \citep{BBS01}.
In the presence of helicity, there is also a turbulent pumping effect,
whose effective velocity is also in the direction of
$H_{\rm K}\overline{\bm{W}}$
\citep{MKTB09}.

In the cases discussed above the turbulence is driven by a helical body
force, which is clearly artificial, but it allows contact to be made with
analytic theories of dynamo action in homogeneous media \citep{Mof78}.
A more realistic case is one where the turbulence is driven by natural
convection in a slab with a temperature gradient in the vertical direction.
Many of the features of dynamo action discussed above carry over to this
case as well, but an additional complication arises both from the fact
that there are impenetrable walls and that the sign of kinetic helicity
changes with depth \citep[e.g.][]{BNPST90,CH06}.

In the present paper we deal with both aspects, but we focus in particular
on the effects of sinusoidal shear, where we expect at least partial
cancellation of the $\alpha$ effect when averaged over horizontal planes.
We contrast our work with earlier results that used linear shear, 
implemented via the shearing-box 
approximation \citep{KKB08}, as well as the case with no shear
\citep{KKB09b}, where only the $\alpha$ effect can operate.
The conclusion from these studies is that in the
simulation domain there is an $\alpha$ effect
of the strength expected from kinematic mean-field theory 
\citep{KKB09a,KKB09b}.
There is also a back-reaction of the magnetic field through the Lorentz force,
and its strength varies depending on whether or not magnetic helicity is
allowed to escape from the domain \citep{KKB09c}.
Again, these aspects are now well understood using mean-field theory.
The new aspect here is the sinusoidal shear.
In a recent paper, \cite{HP09} present results from
convection simulations with rotation and large-scale shear and report
the emergence of a large-scale magnetic field whose growth rate is proportional
to the shear rate, similar to the earlier results of 
\cite{KKB08}.
They also determine the $\alpha$ effect from
their simulations using the so-called imposed-field method and find
that $\alpha$ is small and unaffected by the presence of shear. From
these results the authors conclude that the dynamo cannot be explained by
a classical $\alpha^2$ or $\alpha \Omega$ dynamo.

The interpretation of the results of \cite{HP09} is potentially
in conflict with that of \cite{KKB08}.
In both cases, convection together with shear was found to produce
large-scale fields, but in \cite{KKB08} they are interpreted as being
the result of a conventional $\alpha$ effect while in \cite{HP09}
it is argued that they are due to another mechanism similar to the
incoherent $\alpha$--shear effect \citep{VB97,Sok97,Sil00,Pro07},
or perhaps the shear--current effect \citep{RK03,RK04}.
Moreover, \cite{HP09} argue that the $\alpha$ effect is ruled out.

At this point we cannot be sure that there is really a difference in
interpretations, because the systems considered by \cite{KKB08}
and \cite{HP09} are different in at least two important aspects.
Firstly, in \cite{HP09} there is no density stratification,
and since
$\alpha$ is supposed to be proportional to the logarithmic density gradient
\citep{SKR66} the resulting $\alpha$ may indeed vanish.
However, due to the impenetrable vertical boundaries, the turbulence 
is inhomogeneous so that $\bm\nabla\ln\urms\neq{\bm0}$, which can 
also lead to an $\alpha$ effect 
\citep[e.g.][]{GZR05}.
Here, $\urms$ is the rms velocity of the turbulence.
Secondly, the shear profile changes sign in the horizontal direction.
Together with the vertical inhomogeneity this also produces an $\alpha$ effect
\citep{RK03,RS06}, but its contribution is not captured by horizontal
averaging and it partially cancels the $\alpha$ effect from rotation.
This should be a measurable effect which was not quantified in \cite{HP09}.
Doing this is one of the main motivations behind our present paper.

There is yet another important issue relevant to determining $\alpha$ in
a system where the magnetic Reynolds number is large enough to result in
dynamo action \citep{HdSKB09}.
Obviously, any successful $\alpha$ effect should produce large-scale
magnetic fields.
Given enough time, this field should reach saturation.
By employing a weak external field one might therefore measure $\alpha$
at a saturated level.
Depending on boundary conditions, which were unfortunately not specified in
\cite{HP09}, the saturation can result in a catastrophically quenched
$\alpha$ effect.
Furthermore, here we show that even in the absence of a dynamo the
electromotive force from long time averages reflects not only $\alpha$ 
due to the uniform imposed field as assumed by \cite{HP09}, but also 
picks up contributions from the additionally generated nonuniform 
fields of comparable magnitude.
These caveats in determining $\alpha$ with an externally imposed field
were known for some time \citep{OSBR02,KKOS06}, but they have only recently
been examined in detail \citep{HdSKB09} and were therefore not addressed
by \cite{HP09}.
This gives another motivation to our study.

Here we use a similar simulation setup as \cite{HP09}
and derive the $\alpha$ effect with the imposed-field method. We show
that the value of $\alpha$ determined by the method of resetting the
magnetic field after regular time intervals yields a substantially higher value than
that reported by \cite{HP09}. Furthermore, we show that for a
sinusoidally varying shear, also the $\alpha$ effect will have a
sinusoidal variation in the horizontal direction,
hence explaining why \cite{HP09} did not see the
contribution of shear in their horizontally averaged results.

\section{The model}

In an effort to compare with the study of \cite{HP09}, we use a
Cartesian domain with $L_x=L_y=5d$ and $L_z=d$ with $0<z<d$, where 
$d$ is the depth
of the convectively unstable layer.
We solve the usual set of hydromagnetic equations
\begin{eqnarray}
\frac{\pd \bm{A}}{\pd t} &=& \bm{U}\times\bm{B} - \eta \mu_0 \bm{J}, \\
\frac{D \ln \rho}{Dt} &=& -\bm\nabla\cdot\bm{U}, \\
\frac{D \bm{U}}{Dt} &=& -\frac{1}{\rho}{\bm \nabla}p + {\bm g} - 2\,\bm{\Omega} \times \bm{U} + \frac{1}{\rho} \bm{J} \times {\bm B} \nonumber \\  && \hspace{1.5cm} +  \frac{1}{\rho} \bm{\nabla} \cdot 2 \nu \rho \mbox{\boldmath ${\sf S}$} + \frac{1}{\tau} (\bm{U}-\meanv{U}^{(0)}),\label{equ:mom}\\
\frac{D e}{Dt}\!&=&\!-\frac{p}{\rho}\bm\nabla\cdot{\bm U} + \frac{1}{\rho} \bm{\nabla} \cdot K \bm{\nabla}T + 2 \nu \mbox{\boldmath ${\sf S}$}^2 + \frac{\mu_0\eta}{\rho} \bm{J}^2,
\end{eqnarray}
where $D/Dt=\pd/\pd t + \bm{U}\cdot\bm\nabla$ is the advective time
derivative, $\bm{A}$ is the
magnetic vector potential, $\bm{B} = \bm{\nabla} \times \bm{A}$ the
magnetic field, and $\bm{J} = \mu_0^{-1} \bm{\nabla} \times \bm{B}$ is
the current density, $\mu_0$ is the vacuum permeability, $\eta$ and
$\nu$ are the magnetic diffusivity and kinematic viscosity,
respectively, $K$ is the heat conductivity, $\rho$ is the density,
$\bm{U}$ is the velocity, $\bm{g} = -g\hat{\bm{z}}$ the gravitational
acceleration, and $\bm{\Omega}=\Omega_0(0,0,1)$ the rotation
vector. The fluid obeys an ideal gas law $p=\rho e (\gamma-1)$, where
$p$ and $e$ are the pressure and internal energy, respectively, and
$\gamma = c_{\rm P}/c_{\rm V} = 5/3$ is the ratio of specific heats at
constant pressure and volume, respectively. The specific internal
energy per unit mass is related to the temperature via $e=c_{\rm V}
T$. The rate of strain tensor $\mbox{\boldmath ${\sf S}$}$ is given by
\begin{equation}
{\sf S}_{ij} = \onehalf (U_{i,j}+U_{j,i}) - \onethird \delta_{ij} \bm\nabla\cdot\bm{U}.
\end{equation}
The last term of \Eq{equ:mom} maintains a shear flow of the form 
\begin{equation}
\meanv{U}^{(0)} = U_0 \cos\left[\frac{2\pi(x-x_0)}{L_x} \right]\hat{\bm{e}}_y,\label{equ:sf}
\end{equation}
where $U_0$ is the amplitude of the shear flow, $x_0=-L_x/2$ is the
position of the left-hand boundary of the domain, and 
$\tau$ is a relaxation time. Here we use a $\tau=20\sqrt{d/g}$ which
corresponds to roughly 3.5 convective turnover times.

In their study, \cite{HP09} use the Boussinesq approximation and thus
neglect density stratification.
Here we use the {\sc Pencil
Code}\footnote{http://pencil-code.googlecode.com} which is
fully compressible.
However, in order to stay close to the setup of \cite{HP09} we
employ a weak stratification: the density difference between the top
and the bottom of the domain is only ten per cent and the average Mach
number is always less than 0.1. Hence the effects of compressibility
are small.
The stratification in the associated hydrostatic
initial state can be described by a polytrope with index $m=1$. Unlike
our previous studies \citep[e.g.][]{KKB08}, no stably stratified
layers are present.

The horizontal boundaries are periodic.
We keep the temperature fixed
at the top and bottom boundaries. For the velocity we apply
impenetrable, stress-free conditions according to 
\begin{eqnarray}
\pd_zU_x = \pd_z U_y = U_z=0.
\end{eqnarray}
For the magnetic field we use vertical field conditions
\begin{eqnarray}
B_x = B_y=0,
\end{eqnarray}
that allow magnetic helicity to escape from the domain.

\begin{table}
\caption{Summary of the runs. Here ${\rm Ma}=U_{\rm rms}/\sqrt{dg}$, 
where $U_{\rm rms}$ is the total rms velocity including the shear flow,
${\rm Ma}_0=\urms/\sqrt{dg}$, 
and $\tilde\brms=\brms/B_{\rm eq}$, where $B_{\rm eq}=\sqrt{\mu_0 \rho \urms^2}$.
We use $\Rem\approx18$, $\Co\approx2.3$, and ${\rm Ra}=10^5$ in 
all runs.}
\vspace{12pt}
\centerline{\begin{tabular}{lccccc}
Run & $\rm Ma$ & ${\rm Ma}/{\rm Ma_0}$ & $\Sh$  & $\tilde\brms$ & Dynamo \\
\hline
A0  &  $0.028$ & $1.00$                & $0.00$ & --            &  no    \\
A1  &  $0.027$ & $0.98$                & $0.07$ & --            &  no    \\
A2  &  $0.028$ & $1.01$                & $0.14$ & $0.70$        &  yes   \\
A3  &  $0.039$ & $1.42$                & $0.36$ & $1.15$        &  yes   \\
A4  &  $0.063$ & $2.28$                & $0.72$ & $1.97$        &  yes   \\
A5  &  $0.096$ & $3.47$                & $1.45$ & $3.99$        &  yes
\label{Runs}\end{tabular}}\end{table}

\subsection{Units, nondimensional quantities, and parameters}
Dimensionless quantities are obtained by setting
\begin{eqnarray}
d = g = \rho_0 = c_{\rm P} = \mu_0 = 1\;,
\end{eqnarray}
where $\rho_0$ is the density at $z_{\rm m}=\onehalf d$. The units of
length, time, velocity, density, specific entropy, and magnetic field are then
\begin{eqnarray}
&& [x] = d\;,\;\; [t] = \sqrt{d/g}\;,\;\; [U]=\sqrt{dg}\;,\;\; \nonumber \\ && [\rho]=\rho_0\;,\;\; [s]=c_{\rm P}\;,\;\; [B]=\sqrt{dg\rho_0\mu_0}\;. 
\end{eqnarray}
The simulations are controlled by the following dimensionless
parameters: thermal and magnetic diffusion in
comparison to viscosity are measured by the Prandtl numbers
\begin{eqnarray}
\Pr=\frac{\nu}{\chi_0}, \quad \Pm=\frac{\nu}{\eta},
\end{eqnarray}
where $\chi_0=K/(c_{\rm P} \rho_0)$ is the reference value of the
thermal diffusion coefficient, measured in the middle of the layer,
$z_{\rm m}$, in the non-convecting initial state. 
We use $\Pr=0.6$ and $\Pm=2$ in most models.
Note that \cite{HP09} use $\Pr=1$ and $\Pm=5$,
but based on earlier parameter studies \citep{KKB09a,KKB09c}
we do not expect this difference to be significant.
The efficiency of
convection is measured by the Rayleigh number
\begin{eqnarray}
\Ra=\frac{g d^4}{\nu \chi_0}\left(- \frac{1}{c_{\rm P}}\frac{{\rm d}s}{{\rm d}z} \right)_{z_{\rm m}},
\end{eqnarray}
again determined from the initial non-convecting state at $z_{\rm m}$. The
entropy gradient can be presented in terms of logarithmic temperature
gradients
\begin{eqnarray}
\left(- \frac{1}{c_{\rm P}}\frac{{\rm d}s}{{\rm d}z} \right)_{z_{\rm m}}=\frac{\nabla-\nabla_{\rm ad}}{H_{\rm P}},
\end{eqnarray}
with $\nabla=(\pd \ln T/\pd \ln p)_{z_{\rm m}}$, $\nabla_{\rm ad}=1-1/\gamma$,
and $H_{\rm P}$ being the pressure scale height at $z=z_{\rm m}$.

The effects of viscosity and magnetic
diffusion are quantified respectively by the fluid and magnetic Reynolds numbers
\begin{eqnarray}
\Rey=\frac{\urms}{\nu \kef}, \quad \Rem=\frac{\urms}{\eta \kef}=\Pm\,\Rey,
\end{eqnarray}
where $\urms$ is the root-mean-square (rms) value of the velocity
taken from a run where $\meanv{U}^{(0)}={\bm0}$, 
and $\kef=2\pi/d$ is the wavenumber
corresponding to the depth of the convectively unstable layer.
The strengths of rotation and shear are measured by the Coriolis and
shear numbers
\begin{eqnarray}
\Co=\frac{2\Omega}{\urms \kef}, \quad \Sh=\frac{S}{\urms \kef},
\end{eqnarray}
where $S=2\pi U_0/L_x$.

The size of error bars is estimated by dividing the time series into
three equally long parts.
The largest deviation of the average for each of the three parts from that
over the full time series is taken to represent the error.

\begin{figure}
\begin{center}
\includegraphics[width=\columnwidth]{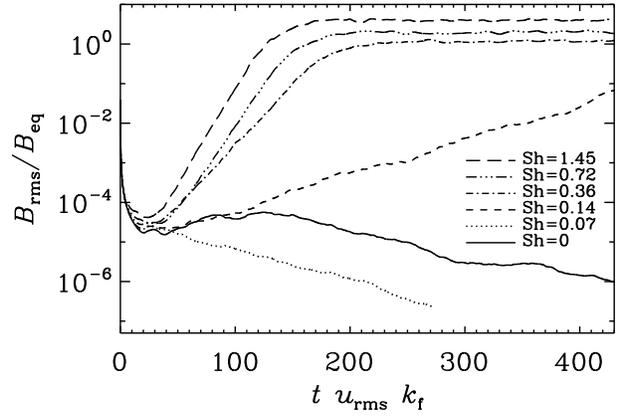}
\end{center}
\caption[]{Root mean square value of the total magnetic field as a
  function of time for the runs listed in Table~\ref{Runs}.}
\label{pbrms}
\end{figure}

\section{Results}

\subsection{Dynamo excitation}
We first set out to reproduce the results of \cite{HP09}. To achieve
this, we take a run with parameters close to theirs which does not
act as a dynamo in the absence of shear ($\Sh=0$). For this baseline
simulation we choose the parameters $\Rem\approx18$ and
$\Co\approx2.3$. We then follow the same procedure as \cite{HP09} and
gradually increase $\Sh$ whilst keeping all other parameters constant
(Table~\ref{Runs}) and determine the growth rate $\lambda$ of the magnetic
field.

\begin{figure}
\begin{center}
\includegraphics[width=\columnwidth]{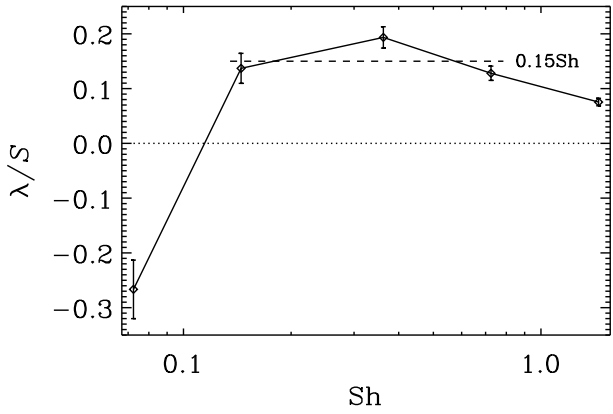}
\end{center}
\caption[]{Growth rate $\lambda$ of the total magnetic field, divided
  by the shear rate $S$ as a function of $\Sh$.}
\label{pgr}
\end{figure}

The time evolution of the rms-value of the total magnetic field from
our set of runs is presented in \Fig{pbrms}. We find no dynamo for
$\Sh=0$ and for weak shear with $\Sh=0.07$, the growth rate of the
field remains virtually the same as in the absence of shear. This
can be understood as follows: imposing large-scale shear via a
relaxation term effectively introduces a friction term for $U_y$ in
places where $\bm{U}-\meanv{U}^{(0)}\neq\bm{0}$, hence lowering the Reynolds
number somewhat. However, as the same relaxation time $\tau \urms
\kef\approx3.5$ is used in all runs with shear, we are confident that
these runs can be compared with each other. As the shear is
increased beyond $\Sh=0.07$, the growth rate first increases roughly directly
proportional to the shear rate $S$ (\Fig{pgr}). However, for
$\Sh>0.72$ the increase of the growth rate slows down similarly as in
several previous studies \citep{Yousef2,KKB08,HP09}.

\begin{figure*}
\begin{center}
\includegraphics[width=\textwidth]{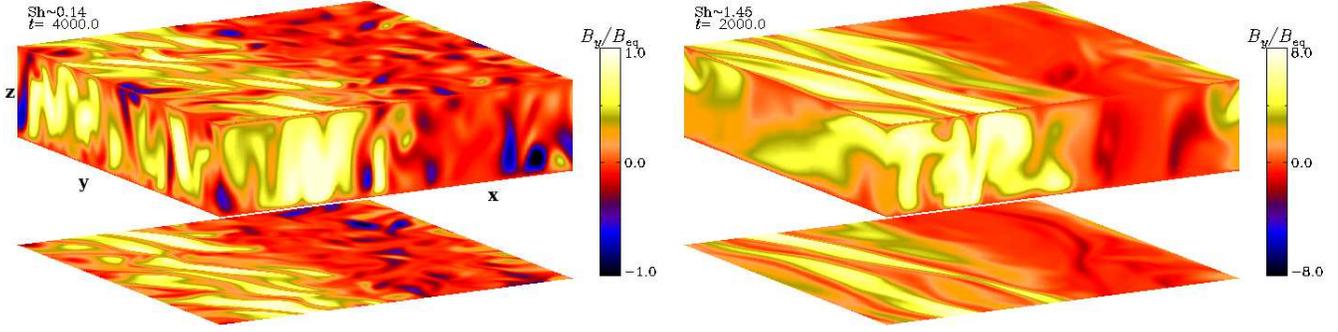}
\end{center}
\caption[]{Magnetic field component $B_y$ in the saturated state from 
  two runs with weak
  (left panel, $\Sh\approx0.14$, $t \urms \kef\approx 700$) and strong 
  shear (right panel,
  $\Sh\approx1.45$, $t \urms \kef\approx 350$). The sides of the 
  boxes show the field at
  the periphery of the domain whereas the bottom (top) panel depicts
  $B_y$ from $z=0.05d$ ($z=0.95d$).}
\label{boxes_By}
\end{figure*}

\subsection{Field structure}
In earlier studies where a homogeneous shear flow was used, the
large-scale magnetic field in the saturated state was non-oscillating,
showed little dependence on horizontal coordinates, and could hence be well
represented by a horizontal average \citep{KKB08}.
However, in the
present case with sinusoidal shear, the field structure and temporal
behaviour can in principle be more complicated. Furthermore, \cite{HP09} do
not comment on the field structure in their study.
In fact, the only evidence of a large-scale field in their paper is
given in the form of spectra of the magnetic field.

\begin{figure}
\begin{center}
\includegraphics[width=\columnwidth]{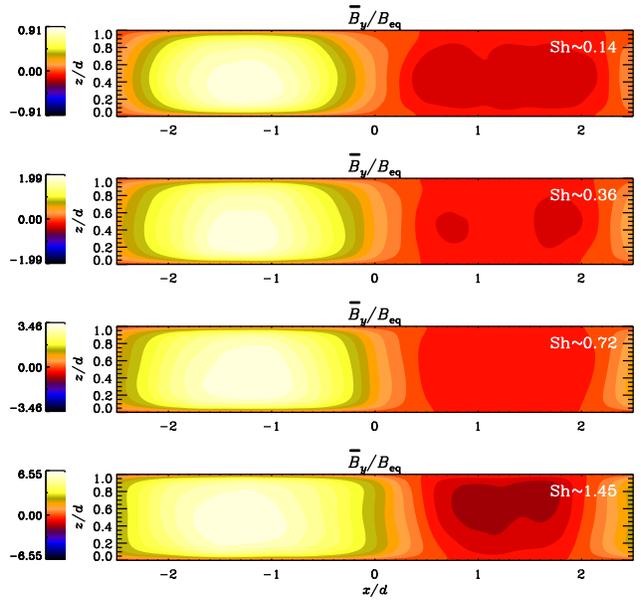}
\end{center}
\caption[]{Magnetic field component $B_y$ averaged over the saturated 
  state in time and over the
  $y$-dimension from Runs~A2-A5.}
\label{pByxz}
\end{figure}

We find that in our simulations the large-scale field is non-oscillating.
It turns out that the magnetic field shows an interesting spatial dependence.
In \Fig{boxes_By} we show visualizations of the structure of the $B_y$
component from the runs with
the weakest ($\Sh\approx0.14$) and the strongest
($\Sh\approx1.45$) shear in which dynamo action was detected.
In both cases it is clear that the strong large-scale fields are
concentrated to one side of the computational
domain whereas the other side of the box is almost devoid of
strong coherent fields. This behaviour is even more striking when the
field is averaged over $y$ and $t$; see \Fig{pByxz}. In the next
section we show that the region of strong large-scale fields
coincides with the region where the $\alpha$ effect is strongest.

\subsection{$\alpha$ effect}
The origin of large-scale magnetic fields in helical turbulence is
commonly attributed to the $\alpha$ effect in turbulent dynamo
theory \citep[e.g.][]{Mof78,KR80,RH04}. Results for convection
simulations, making use of the test-field method \citep{KKB09b},
suggest that the $\alpha$ effect does
indeed contribute to large-scale dynamo action in simulations
presented by \cite{KKB08}.
However, it was also shown that, in order to fully explain the
simulation results, additional contributions from the shear--current
and $\meanv{\Omega}\times \meanv{J}$ effects \citep{R69} appear 
to be needed.

On the other hand, \cite{HP09} claim that in their setup the $\alpha$ effect
is small, unaffected by shear, and thus incapable of driving a
large-scale dynamo. The setup of \cite{HP09} is based on the Boussinesq
approximation whereby stratification is not present in their
system. However, the impenetrable vertical boundaries also generate an
inhomogeneity, which, in a rotating system leads to an $\alpha$ effect
of the form \citep{SKR66}
\begin{equation}
\alpha_{ij}^{(\Omega)}=\alpha_1 (\bm{G}\cdot\bm\Omega)\delta_{ij} + \alpha_2 (G_i\Omega_j + G_j\Omega_i),
\end{equation}
where $G_i$ denotes the inhomogeneity and $\bm\Omega$ is the rotation
vector. In Boussinesq convection with rotation the kinetic helicity
and thus the $\alpha$ effect are antisymmetric around the midplane of the
layer. In such cases it can be useful to average over one vertical half of the
layer to obtain an estimate of $\alpha$. We note that mean-field
dynamo models have shown that the details of the $\alpha$ profile 
can also play a significant role \citep[e.g.][]{BS87,SG03}.
In what follows, we show in most cases the full profile of
$\alpha$ and present averages over the upper half of the domain only when comparing
directly to \cite{HP09}. 
Since the simulations in the present paper are weakly
stratified, only minor deviations from a perfectly symmetric profile can
be expected to occur.

Adding a shear flow of the form presented in \Eq{equ:sf} produces
large-scale vorticity $\mean{W}_z\propto\sin \tilde{x}$,
where $\tilde{x}$ is a shifted and rescaled $x$ coordinate
with $\tilde{x}=2\pi(x-x_0)/L_x$.
Such vorticity leads to an $\alpha$ effect
\citep[see, e.g.][]{RK03,RS06},
\begin{equation}
\alpha_{ij}^{(\meanv{W})}=\alpha_1 (\bm{G}\cdot\meanv{W})\delta_{ij} + \alpha_2 (G_i\mean{W}_j + G_j\mean{W}_i),\label{alpW}
\end{equation}
which, in the present case, leads to $\alpha_{yy}\propto\sin \tilde{x}$.
Thus, when both rotation and shear are present, $\alpha=\alpha(x,z)$ is a
function of both $x$ and $z$.

\begin{figure}
\begin{center}
\includegraphics[width=\columnwidth]{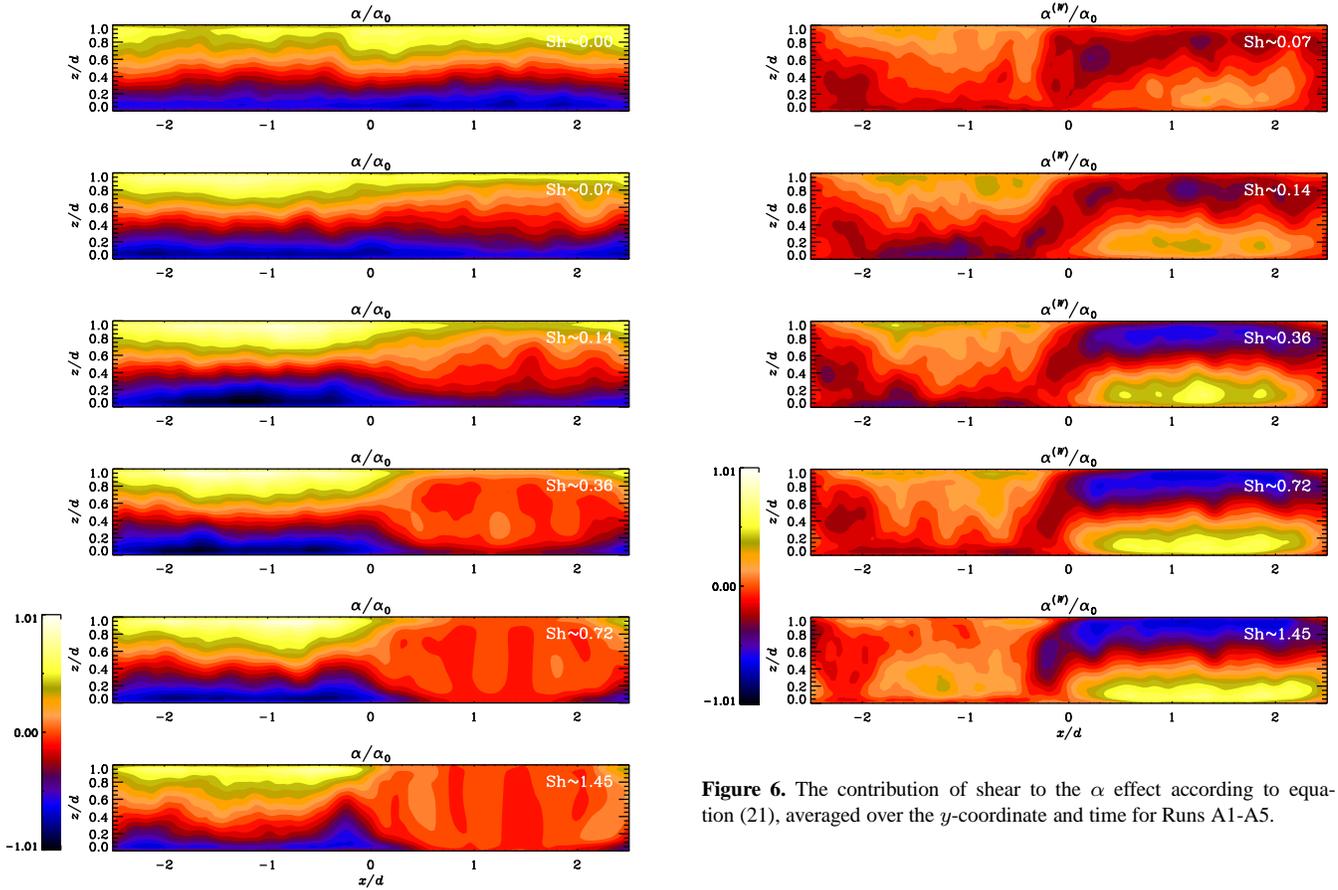}
\end{center}
\caption[]{The coefficient $\alpha$, averaged over the $y$-direction
  and time for Runs~A0-A5.}
\label{palpyy}
\end{figure}

\begin{figure}
\begin{center}
\includegraphics[width=\columnwidth]{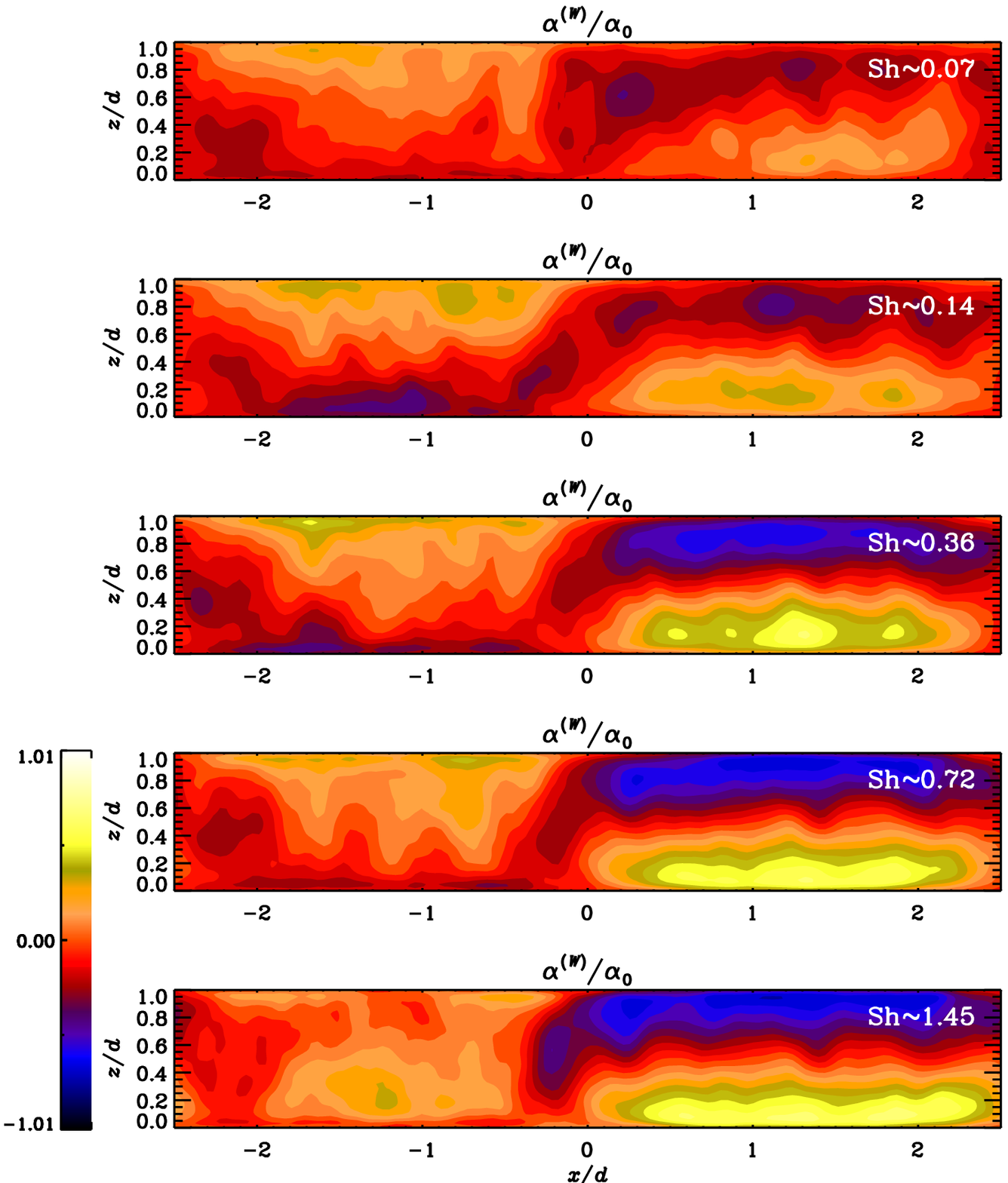}
\end{center}
\caption[]{The contribution of shear to the $\alpha$ effect according
  to \Eq{shearalp}, averaged over the $y$-coordinate and time for
  Runs~A1-A5.}
\label{palpyy_shear}
\end{figure}

In order to measure the $\alpha$ effect, we impose a weak uniform
magnetic field $B_0\hat{\bm{e}}_y$, with $B_0\approx4\cdot10^{-5} B_{\rm
  eq}$, and measure the response of the relevant ($y$)
component of the electromotive force. Our $\alpha$ is then obtained
from
\begin{equation}
\alpha \equiv \alpha_{yy} = \mathcal{\mean{E}}_{y}/B_0.\label{equ:alpha}
\end{equation}
In contrast to the study of \cite{HP09}, we do not usually allow the field
that is generated in addition to $B_0$ to saturate, but reset it after
a time interval $\Delta t \approx 10\ t \urms \kef$. Such a procedure
was first introduced by \cite{OSBR02} and it was used also in \cite{KKOS06}
to circumvent the complications that arise due to the additionally
generated fields. A more systematic study of \cite{HdSKB09} showed
that only if $\Delta t$ is not too long, the kinematic value of
$\alpha$ can be obtained
if there is a successful large-scale dynamo present in the system. 
However, in the present study and also in that of \cite{HP09} there is no 
dynamo in the runs from which $\alpha$ is computed. We find that it is 
still necessary to use resetting to obtain the correct value of 
$\alpha$ even in the absence of a dynamo. However, we postpone detailed 
discussion of this issue to Section \ref{ImportanceOfResetting}.

\begin{figure}
\begin{center}
\includegraphics[width=\columnwidth]{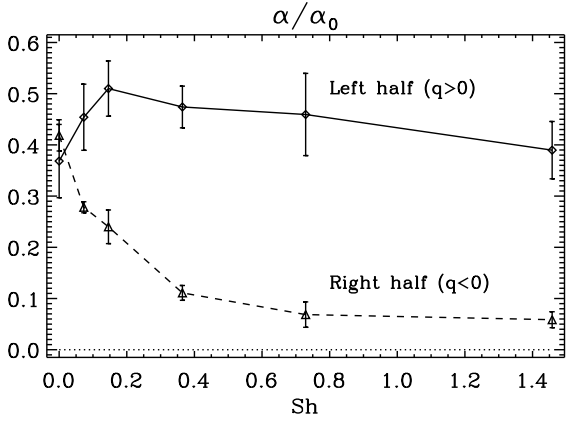}
\end{center}
\caption[]{The coefficient $\alpha$ averaged over the upper half
  ($0.5d<z<d$) of the domain from the left ($x<0$, solid line) and
  right ($x>0$, dashed line) sides of the box.}
\label{palpside}
\end{figure}

Our results for $\alpha$ from runs with constant rotation and varying
shear are shown in \Fig{palpyy}. We find that in the absence of shear,
$\alpha$ is a function only of $z$ and has a magnitude of about
$0.6\alpha_0$,
where $\alpha_0=\onethird \urms$ is a reference value, and $\urms$ 
is taken from a run with
$\Sh=0$. When shear is introduced, $\alpha$ increases
(decreases) in the regions of the domain where $\sin \tilde{x}>0$ ($\sin
\tilde{x}<0$). However, for strong shear, the contribution to $\alpha$ from
shear no longer appears to be symmetric around $x=0$. This can be understood in
terms of the shear parameter
\begin{equation}
q=-\frac{\pd \mean{U}^{(0)}}{\pd x}/\Omega,
\end{equation}
where
\begin{equation}
\frac{\pd \mean{U}^{(0)}}{\pd x}=S\sin\tilde{x}.
\end{equation}
The flow is linearly unstable for $q>2$ (Rayleigh instability criterion).
Although the maximum value of $q$ in our
simulations is about $1.25$ it is clear that for $\Sh\ga0.36$ (with $|q|\ga0.31$), the
profile and the magnitude of $\alpha$ are no longer significantly affected by
the increasing shear. 
In order to illustrate this we compute the contribution of $\alpha$
due to shear from runs with $\Sh\neq0$ by subtracting the
$\alpha$ that was found in the absence of shear using
\begin{equation}
\alpha^{(\mean{W})}=\alpha-\alpha^{(\Omega)},\label{shearalp}
\end{equation}
where $\alpha^{(\Omega)}$ is the $\alpha$ obtained from Run~A0 with no
shear but only rotation.
The results are shown in \Fig{palpyy_shear} and clearly show that for
small $\Sh$ ($\la0.14$), the shear-induced $\alpha$ shows a
sinusoidal variation as a function of $x$. 
For larger shear the profile of $\alpha^{(\mean{W})}$ is no longer 
antisymmetric around
$x=0$. This could reflect the asymmetry of the results for $q>0$
($-L_x/2<x<0$) and $q<0$ ($0<x<L_x/2$), that was found earlier
by Snellman et al.\ (2009) in a somewhat
different context of forced turbulence under the influence
of rotation and shear.
They found that the 
Reynolds stresses were significantly different in setups with
different sign of $\Sh$ or $q$, and that this asymmetry became more
pronounced when the magnitude of shear was increased. Similar behavior
has been seen in the magnetohydrodynamic regime by Korpi et
al.\ (2009) in the Reynolds and Maxwell stresses. 

We also observe that the magnitude of $\alpha^{(\mean{W})}$ does not
significantly change for $\Sh\ga0.36$.
This could indicate that the
$\alpha$ effect due to shear saturates and that a simple relation like
\Eq{alpW} is no longer valid. 
This is apparent from
Figure~\ref{palpside} which shows $\alpha$ volume-averaged over the
upper half of the domain separately for the left and right sides of the 
box. For weak shear ($\Sh\la0.2$) we find that $\alpha$ is linearly
proportional to shear. For $\Sh\la0.4$ the values of $\alpha$ on both 
sides appear to saturate to constant values. The results thus imply that 
the coefficients $\alpha_1$ and $\alpha_2$ in \Eq{alpW} should  
depend on $\meanv{W}$ when shear is strong.
We note that in \cite{HP09} also larger
values of shear were used.
The large vortex seen in the velocity
field in their Figure~3 indicates that some of their runs with strong
shear could indeed be in the Rayleigh-unstable regime.

With the present data we cannot ascribe the appearance of the
large-scale dynamo solely to the $\alpha$ effect. However, the
coincidence of regions of strong magnetic fields and large $\alpha$
suggest that the $\alpha$ effect is indeed an important ingredient in
generating the large-scale fields.

\begin{table}
\caption{Summary of runs with and without resetting with
varying $B_0$. Run~B1 corresponds to Run~A0 in 
Table~\ref{Runs}. $\Co\approx2.3$, $\Sh=0$, and ${\rm Ra}=10^5$
in all runs and the imposed field in normalised form is given by 
$\tilde{B}_0=B_0/\Beq$.}
\vspace{12pt}
\centerline{\begin{tabular}{lcccccc}
Run & $\Rem$ & $\tilde{B}_0$  & $\mean{\alpha}/\alpha_0$ & Resetting \\
\hline
B1  &  $18$ & $4\cdot10^{-5}$ & $0.39\pm0.05$            &  yes   \\
B2  &  $18$ & $0.04$         & $0.36\pm0.03$            &  yes   \\
B3  &  $18$ & $0.11$         & $0.37\pm0.05$            &  yes   \\
B4  &  $18$ & $0.39$         & $0.63\pm0.21$            &  yes   \\
B5  &  $18$ & $1.25$         & $0.25\pm0.10$            &  yes   \\
B6  &  $18$ & $4.47$         & $0.06\pm0.05$            &  yes   \\
\hline
C1  &  $18$ & $4\cdot10^{-5}$ & $0.09\pm0.06$            &  no   \\
C2  &  $18$ & $0.04$         & $0.12\pm0.09$            &  no   \\
C3  &  $18$ & $0.12$         & $0.09\pm0.02$            &  no   \\
C4  &  $18$ & $0.37$         & $0.12\pm0.05$            &  no   \\
C5  &  $18$ & $1.27$         & $0.08\pm0.03$            &  no   \\
C6  &  $18$ & $2.22$         & $0.06\pm0.01$            &  no   \\
C7  &  $18$ & $4.10$         & $(1.10\pm0.34)\cdot10^{-3}$ &  no   \\
\hline
D1  &  $30$ & $4\cdot10^{-5}$ & $0.36\pm0.03$            &  yes  \\
D2  &  $30$ & $4\cdot10^{-5}$ & $-0.03\pm0.23$           &  no
\label{Resetruns}\end{tabular}}\end{table}

\subsection{Importance of resetting}
\label{ImportanceOfResetting}

It has previously been demonstrated that the imposed field 
method can yield misleading results if a successful large-scale dynamo
is operating in the system and long time averages are 
employed \citep{HdSKB09}.
In this case, unexpectedly low values of $\alpha$ could be explained
by the fact that the system is already in a saturated state.
However, many papers have reported small
values of $\alpha$ also for systems that do not act as 
dynamos \citep[e.g.][]{CH06,HC08,HP09}.
These results in apparent contradiction with 
those of \cite{OSBR02,KKOS06,KKB09a} who use either the imposed field
method with resetting or the test field method.
In these cases the systems must be in a truly kinematic state.
Thus, the explanation of \cite{HdSKB09} does not apply.
The purpose of this section is therefore to resolve this puzzle.

\begin{figure}
\begin{center}
\includegraphics[width=\columnwidth]{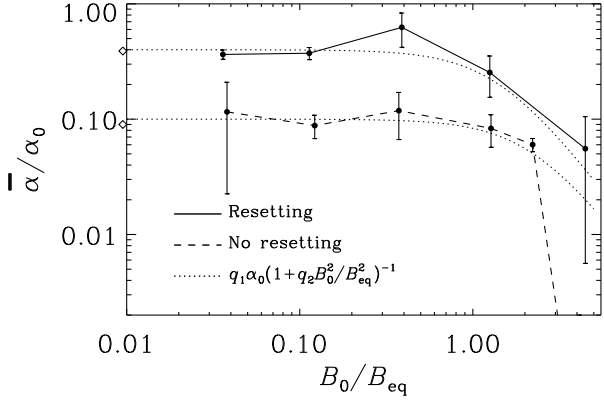}
\end{center}
\caption[]{Coefficient $\mean\alpha$  according
  to \Eq{altop} as a function of $B_0$ from
  runs where the field is either being reset (Set~B, solid line) or
  left to evolve on its own (Set~C, dashed line). The dotted lines
  show fits to a quenching formula given in the legend where we use
  the coefficients $q_1=0.4$ ($0.1$) and $q_2=0.5$ ($0.2$) in the
  upper (lower) curve. The diamonds on the left of the vertical axis
  indicate the values of $\mean\alpha$ for
  $B_0/\Beq\approx4\cdot10^{-5}$.}
\label{palp_B0}
\end{figure}

We begin the investigation of this issue by performing two sets of 
simulations where we study the dependence of $\alpha$, as measured 
using \Eq{equ:alpha}, on $B_0$ with runs where the field is being 
periodically reset or left to evolve unimpeded (Sets~B and C,
see Table~\ref{Resetruns}). We take 
Run~A0 with $\Rem\approx18$ and no shear as our baseline and vary
$B_0/B_{\rm eq}$ in the range $4\cdot10^{-5}\ldots4$. Our results for 
$\mean\alpha$, defined as the volume average over the upper half of 
the box,
\begin{equation}
\mean\alpha=\frac{2}{L_z}\int_{\onehalf L_z}^{L_z} \frac{\memfy(z)}{B_0} dz,\label{altop}
\end{equation}
are shown in \Fig{palp_B0}. We see that, with the exception of the
strongest $B_0$ case in Set C, the
results for both sets are in accordance with a simple quenching
formula 
\begin{equation}
\mean\alpha=\frac{q_1 \alpha_0}{1+q_2(B_0/\Beq)^2},
\end{equation}
where $q_1$ and $q_2$ are constants which we use as free parameters in
the fitting. We find that the value of $\mean\alpha$ for weak fields is
consistently four times smaller in the cases where no resetting is
performed. The values of $\mean\alpha$ in the range 
$B_0/\Beq\approx0.04\ldots1$ are essentially the same as those made 
for our standard imposed field strength 
$B_0/\Beq\approx4\cdot10^{-5}$ (see also Table~\ref{Resetruns}). This 
suggests that the values of $\mean\alpha$ in this range represent the 
kinematic stage and that the factor of four between the results in 
the different sets arises from the additional inhomogeneous mean magnetic 
fields generated in the cases where no resetting is performed.

\begin{figure}
\begin{center}
\includegraphics[width=\columnwidth]{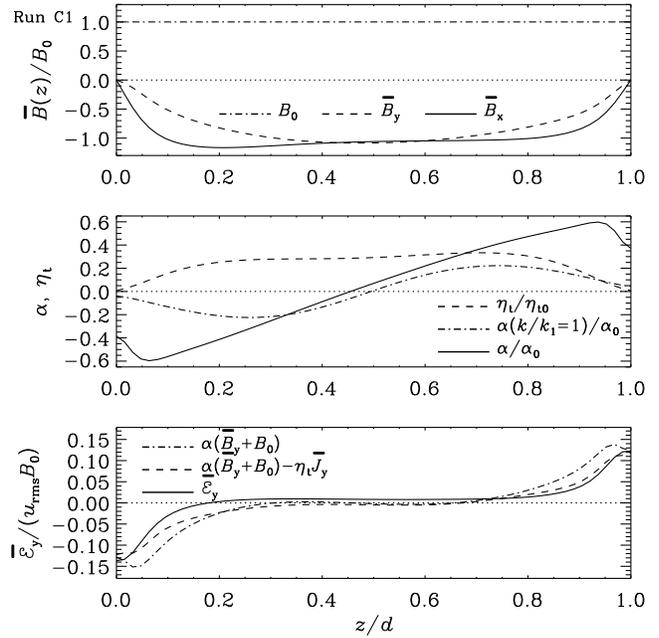}
\end{center}
\caption[]{Top panel: horizontally averaged horizontal components
  of the magnetic field from Run~C1. Middle panel: vertical profiles
  of $\alpha(z)$ from the imposed-field method (solid line) and test 
  field calculation with $k=k_1$ (dash-dotted line), and $\etat(z)$ 
  (dashed line). Bottom panel:
  $y$-component of the electromotive force (solid line) compared with
  $\alpha \mean{B}_y-\etat \mean{J}_y$ (dashed line), and $\alpha
  \mean{B}_y$ (dash-dotted line).}
\label{peymz}
\end{figure}

This is demonstrated in the uppermost panel of \Fig{peymz} where the 
additionally generated horizontal magnetic fields, averaged over time 
and horizontal directions, are shown from Run~C1. The origin of these
fields can be understood as follows: the imposed field 
$B_0 \hat{\bm e}_y$ induces a $z$-dependent
electromotive force in the $y$ direction, i.e.\ $\memfy(z)$. 
This leads to the generation of an $x$ component of mean magnetic
field via $\dot{\mean{B}}_x(z)=\ldots-\memf_{y,z}$ which, on the other 
hand, induces a $z$ dependent electromotive force $\memf_x(z)$ and hence
$\dot{\mean{B}}_y(z)=\ldots+\memf_{x,z}$. Since these additional fields
are functions of $z$, mean currents $\mean{J}_x(z)=-\mean{B}_{y,z}$ and 
$\mean{J}_y(z)=\mean{B}_{x,z}$ are also present. We emphasize that these
fields arise due to the presence of an imposed field and decay if 
the imposed field is removed.

It is now clear that $\alpha$ cannot be determined using 
\Eq{equ:alpha} in this situation because the electromotive force picks
up contributions from the generated fields according to
\begin{equation}
\memfy(z)=\alpha(z)[\mean{B}_y(z)+B_0]-\etat(z) \mean{J}_y(z).\label{equ:memfy}
\end{equation}
Here we omit the off-diagonal components of $\alpha_{ij}$ and 
$\eta_{ijk}$ whose influence on the final result is marginal.
Since the magnetic fields are weak, $\alpha$ and $\etat$ can be 
considered as the kinematic values. We use here $\alpha$ as
determined from Run~B1 (imposed field with resetting) and $\etat$ 
obtained from a corresponding test field simulation (see the 
middle panel of \Fig{peymz}) where the test 
fields have a $\sin kz$ dependence on $z$ with $k/k_1=1$ and 
$k_1=2\pi/d$.
For more details about the test field method in the context
of convection simulations see \cite{KKB09a}.
We normalise the turbulent diffusion with a reference value 
$\eta_{\rm t0}=\onethird \urms \kef^{-1}$.
The bottom panel of \Fig{peymz} shows that \Eq{equ:memfy} with these
$z$-dependent
coefficients gives a good fit to the simulation
data of $\memfy$ from Run~C1 when the actual mean magnetic fields
are used.
The diffusion term in \Eq{equ:memfy} has a noticeable effect only 
near the boundaries where the current is also largest. These results
demonstrate that the interpretation of the electromotive force in 
terms of \Eq{equ:alpha} is insufficient if long time averages 
are used.

A general comment is here in order.
Near boundaries, as well as elsewhere in the domain where
the scale of variation of the mean field becomes comparable
with the scale of the turbulent eddies, a simple multiplication
with turbulent transport coefficients becomes inaccurate and one
needs to resort to a convolution with integral kernels.
The kernels can be obtained via Fourier transformation using the
test-field results for different wavenumbers \citep{BRS08}.
In the present paper we have only considered the result for the
wavenumber $k=2\pi/L_z$.
This is also the case for the $\etat$ shown in the middle panel of
\Fig{peymz}.
The $\alpha$ obtained from the test-field method has a more nearly
sinusoidal shape, but with similar amplitude than the profile
shown in \Fig{peymz}.
This confirms the internal consistency of our result.

\begin{figure}
\begin{center}
\includegraphics[width=\columnwidth]{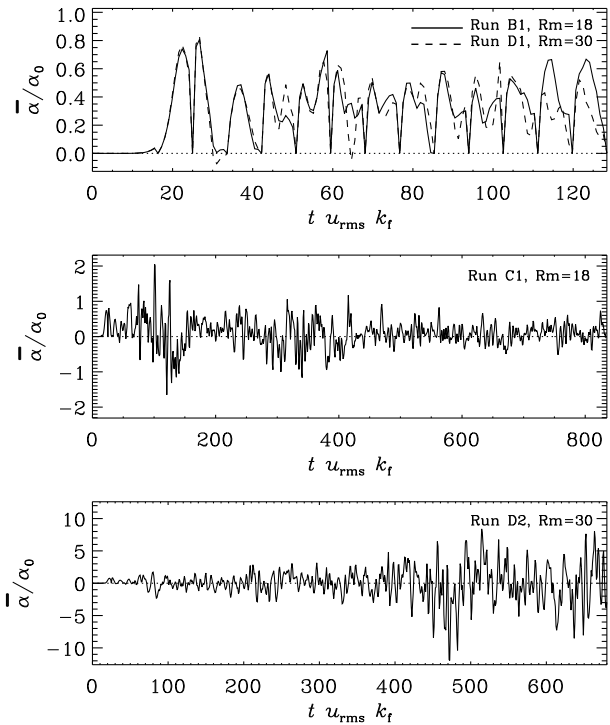}
\end{center}
\caption[]{Time series of the coefficient $\mean\alpha$ for 
  Runs~B1 and D1 (uppermost panel), C1 (middle), and D2 (bottom).}
\label{palp_reset}
\end{figure}

Another facet of the issue is highlighted when the magnetic Reynolds 
number is increased from 18 to 30 (Runs D1 and D2, see \Fig{palp_reset}). 
The larger 
$\Rem$ value is very close to marginal for dynamo action whereas the smaller
value is clearly subcritical. We find that, if
resetting is used, the kinematic value of $\mean\alpha$ is independent 
of $\Rem$ in accordance with mean-field theory.
The situation changes dramatically if we let the field evolve without
resetting; see the two lower panels of \Fig{palp_reset}. For
Run~C1 with $\Rem\approx18$ we can still extract a statistically
significant mean value of $\mean\alpha$ although the scatter of the
data is considerable. For Run~D2 with $\Rem\approx30$ the 
fluctuations of $\mean\alpha$ increase even further so that a
very long time average would be needed to obtain a statistically
meaningful value. A similar convergence issue has been 
encountered in the studies by \cite{CH06,HC08,HP09}. However, as we 
have shown above, the interpretation of such values cannot be done
without taking into account the additionally generated fields and the
effects of turbulent diffusion.

\section{Conclusions}

We use three-dimensional simulations of weakly stratified turbulent
convection with sinusoidal shear to study dynamo action. The
parameters of the simulations are chosen so that in the absence of
shear no dynamo is present. For weak shear the growth rate of the
magnetic field is roughly proportional to the shear rate.
This is in accordance
with earlier studies. A large-scale magnetic field is found in all
cases where a dynamo is excited. The strongest large-scale fields are
concentrated in one half of the domain ($x<0$), with a sign change
close to $x=0$ and weaker field of opposite sign in the other half
($x>0$) of the box.

In an earlier study, \cite{HP09} investigated a similar system and came to
the conclusion that the dynamo cannot be explained by $\alpha \Omega$
or $\alpha^2$ dynamos due to a low value of $\alpha$ determined using
the imposed-field method. 
However, we demonstrate that their method where long time averages 
are used yields the kinematic value $\alpha$ only if
additionally generated inhomogeneous mean fields are taken into 
account. Hence, this analysis becomes meaningless without the 
knowledge of turbulent diffusion.
The situation has now changed through the widespread usage of the test-field method
to obtain values of $\etat$ at the same time \citep[see, e.g.,][]{Gre08}.
Furthermore, we show that, if the
magnetic field is reset before the additionally generated fields 
become comparable to the imposed field, the kinematic value of $\alpha$ 
can be obtained by much shorter simulations and without the 
complications related to gradients of $\meanv{B}$ or statistical 
convergence. These issues were already known for some time 
\citep{OSBR02,KKOS06}, but they have generally not been taken into
consideration.

Another new aspect is the sinusoidal shear that is expected to lead to
a sinusoidal $\alpha$ profile \cite[e.g.][]{RS06}. In the
study of \cite{HP09} a volume average of $\alpha$ over one vertical half of the domain
is used, which averages out the contribution of $\alpha$ due to shear.
We find that, in the absence of shear, $\alpha$ is approximately
antisymmetric with respect to the midplane of the convectively
unstable layer suggesting that the main contribution to $\alpha$ comes
from the inhomogeneity due to the boundaries rather than due to
density stratification. When sinusoidal shear is introduced into the
system, an additional sinusoidal variation of $\alpha$
in the $x$ direction is indeed present. When
the shear is strong enough, the $\alpha$ profile is highly
anisotropic. The maximum value of $\alpha$ is close to the expected one,
$\alpha_0=\onethird\urms$,
which is significantly higher than the $\alpha$ in \cite{HP09}.

We also note that the regions of strong large-scale magnetic fields
coincide with the regions where the $\alpha$ effect is the
strongest.
This supports the idea that the $\alpha$
effect does indeed play a significant role in generating the
large-scale field.

\section*{Acknowledgments}
The authors acknowledge Matthias Rheinhardt for pointing out the importance
of turbulent diffusion in connection with non-uniform mean fields when
no resetting is used.
The numerical simulations were performed with the supercomputers
hosted by CSC -- IT Center for Science in Espoo, Finland, who are
administered by the Finnish Ministry of Education. Financial support
from the Academy of Finland grant Nos.\ 121431 (PJK) and 112020 (MJK),
the Swedish Research Council grant 621-2007-4064, and
the European Research Council under the AstroDyn Research Project 227952 are
acknowledged.
The authors acknowledge the hospitality of NORDITA during the program
``Solar and Stellar Dynamos and Cycles''.

\end{document}